\documentclass[sigconf]{acmart}
\usepackage{threeparttable, tablefootnote}
\usepackage{pifont} % for the colored check and cross
\usepackage{enumitem} % for enumerate resume (continuous sequence number). Usage: `\begin{enumerate}[resume]`
\usepackage{algorithm}
\usepackage{algpseudocode}
\usepackage{amsmath, amsthm}
\usepackage{amsfonts}
\usepackage{breakurl} % For long URLs in cite
\AtBeginDocument{%
  \providecommand\BibTeX{{%
    \normalfont B\kern-0.5em{\scshape i\kern-0.25em b}\kern-0.8em\TeX}}}

\begin{document}
\title{A Multi-Party, Multi-Blockchain Atomic Swap Protocol with Universal Adaptor Secret}

\author{Shengewei You}
\affiliation{%
  \institution{University of Notre Dame}
  % \streetaddress{}
  \city{South Bend}
  \state{Indiana}
  \country{USA}
  \postcode{46556}
}
\email{syou@nd.edu}
\orcid{0003-3156-1372}

\author{Aditya Joshi}
\affiliation{%
  \institution{University of Notre Dame}
  % \streetaddress{}
  \city{South Bend}
  \state{Indiana}
  \country{USA}
  \postcode{46556}
}
% \email{}
% \orcid{}

\author{Andrey Kuehlkamp}
\affiliation{%
  \institution{University of Notre Dame}
  % \streetaddress{}
  \city{South Bend}
  \state{Indiana}
  \country{USA}
  \postcode{46556}
}
% \email{}
% \orcid{}

\author{Jarek Nabrzyski}
\affiliation{%
  \institution{University of Notre Dame}
  \city{South Bend}
  \state{Indiana}
  \country{USA}
  \postcode{46556}
}
% \email{}
% \orcid{}

% %%
% %% By default, the full list of authors will be used in the page
% %% headers. Often, this list is too long, and will overlap
% %% other information printed in the page headers. This command allows
% %% the author to define a more concise list
% %% of authors' names for this purpose.
% \renewcommand{\shortauthors}{You, et al.}


\begin{abstract}
The increasing complexity of digital asset transactions across multiple blockchains necessitates a robust atomic swap protocol that can securely handle more than two participants. Traditional atomic swap protocols, including those based on adaptor signatures, are vulnerable to malicious dropout attacks, which break atomicity and compromise the security of the transaction. This paper presents a novel multi-party atomic swap protocol that operates almost entirely off-chain, requiring only a single on-chain transaction for finalization. Our protocol leverages Schnorr-like signature verification and a universal adaptor secret to ensure atomicity and scalability across any number of participants and blockchains without the need for smart contracts or trusted third parties. By addressing key challenges such as collusion attacks and malicious dropouts, our protocol significantly enhances the security and efficiency of multi-party atomic swaps. Our contributions include the first scalable, fully off-chain protocol for atomic swaps involving any number of participants, adding zero overhead to native blockchains, and providing a practical and cost-effective solution for decentralized asset exchanges.

\end{abstract}





%   \includegraphics[width=\textwidth]{src/images/bridgeProtocol/P3_B1_B2_Intereactions.pdf}
%   \caption{Protocol Overview}
%   \Description{Robust Cross-blockchain Bridge with Autonomous AI Agent Oracle}
%   \label{fig:teaser-protocol-overview}


\maketitle

\section{Introduction}

Blockchain technology has revolutionized the landscape of digital asset transactions, providing a decentralized and secure platform for peer-to-peer exchanges \cite{ren2023interoperability}. Atomic swap protocols have emerged as a pivotal mechanism within this ecosystem, facilitating direct asset exchanges between parties without the need for trusted intermediaries \cite{poelstra2017scriptless,poelstra2017adaptor}. However, the growing complexity of the blockchain ecosystem necessitates the development of efficient, secure, and scalable multi-party atomic swap protocols.

Existing atomic swap protocols, while effective in simple one-to-one exchanges \cite{zhu2024atomic,riahi2024bitcoin}, often struggle to address the challenges posed by multi-party scenarios. These protocols are vulnerable to malicious dropout attacks, where a participant can disrupt the atomicity of the swap, leaving others unable to complete their transactions \cite{crosschainRisk,wang2021sok}. Furthermore, the reliance on multiple on-chain transactions in traditional protocols leads to high costs and inefficiencies, hindering their scalability and practical adoption in complex scenarios \cite{pillai2021burn,lee2023sok}.

Addressing these challenges is paramount for the advancement of decentralized finance (DeFi) and cross-chain interoperability. A robust multi-party atomic swap protocol would not only enhance security and reduce transaction costs but also unlock a wider range of sophisticated financial transactions, fostering a more interconnected and efficient blockchain ecosystem.

In this paper, we propose a novel multi-party atomic swap protocol designed to overcome the limitations of existing approaches. Our protocol leverages advanced cryptographic techniques, including adaptor signatures signatures \cite{poelstra2017scriptless,poelstra2017adaptor} and cryptographic accumulators \cite{benaloh1993one,camenisch2002dynamic,ozdemir2020scaling} to achieve enhanced security, scalability, and efficiency. Please refer to the Appendix for a more formal definition of the ECDSA-based adaptor signature and cryptographic accumulators.

\subsection{Research Problem}

The primary research problem we address is the lack of scalable and secure multi-party atomic swap protocols that can resist malicious dropout attacks and minimize on-chain operations. Specifically, we aim to design a protocol that:

\begin{itemize}
\item Operates predominantly off-chain, requiring only a single on-chain transaction for finalization.
\item Employs a universal adaptor secret to guarantee atomicity even in the presence of malicious dropouts.
\item Supports an arbitrary number of participants and blockchains implementing Schnorr-like signature verification, without relying on smart contracts.
\item Leverages cryptographic accumulators and public key verification for scalability and efficiency.
\item Addresses potential security vulnerabilities, including collusion and impersonation attacks.
\end{itemize}

\subsection{Contributions}

The contributions of this paper are as follows:

\begin{itemize}
\item We introduce a novel multi-party atomic swap protocol that operates primarily off-chain, significantly reducing transaction costs and enhancing efficiency.
\item We propose the concept of a universal adaptor secret, ensuring the atomicity of the swap even when faced with malicious dropout attempts.
\item We demonstrate the scalability of our protocol, supporting any number of participants and blockchains with Schnorr-like signature verification.
\item We present a comprehensive security analysis, addressing potential attack vectors and demonstrating the robustness of our protocol.
% \item We evaluate the performance of our protocol through theoretical analysis and empirical results, showcasing its efficiency and scalability.
\end{itemize}

\section{Related Work and Background}

In this section, we summarize the existing approaches to exchange assets between multiple blockchains. We compare the security and privacy characteristics of our burn-and-mint protocol with each approach and summarize the limitations.
\label{sec:related-works}

    % % Previous bridge
    % \input{src/components/table/table_related_works_bridge_compare}

    %% Updated Version

    \subsection{Atomic Swap}
        Blockchain interoperability protocols aim to facilitate seamless and efficient transactions across distinct blockchain networks. One prominent method to achieve this interoperability is the Hashed Time Lock Contracts (HTLC) that supports atomic swaps, eliminating the need for a trusted third party. The concept of HTLCs was first introduced as a part of the Bitcoin Lightning Network white paper \cite{poon2016bitcoin} by Joseph Poon and Thaddeus Dryja. 
    
        Atomic swap is a method that two parties can use to directly exchange cryptocurrencies across multiple blockchains. A subsequent protocol variation models cross-chain swaps as a directed graph, extending HTLC to an atomic cross-chain swap protocol capable of dealing with strongly-connected digraph swaps \cite{herlihy2018atomic}. However, this approach can incentivize profiteering that leads to asset price fluctuations, causing potential swap declinations. 

        Further research has addressed these challenges through novel mechanisms. For example, Han et al. \cite{han2019optionality} treated atomic swap as an American-style call-option \footnote{The American Call Option is a financial contract that gives an investor the right, but not the obligation, to buy an asset at a specified price before or at a specific future date, which is similar to the atomic swap in cryptocurrency as it offers a choice to transact (swap) assets at agreed terms within a certain timeframe, without mandating it.} without a premium \footnote{A Premium is an upfront cost—that investors pay and must be factored into the overall profitability of the trade.}, proposing a new mechanism with a premium to incentivize fair swaps. Similarly, Heilman et al. \cite{heilman2020arwen} proposed a layer-two-based Request for Quote (RFQ) trading protocol to incentivize swift coin unlocks, thus curtailing lockup griefing. Xue et al. \cite{xue2021hedging} introduced a premium distribution phase into the HTLC to reduce sore loser attacks \footnote{"Sore loser attacks," also known as lockup griefing, refer to a type of strategy in atomic swap transactions where one party intentionally refuses to complete the transaction after it has begun, causing the other party's funds to remain locked up in the contract for the duration of the time lock.}. A more recent development, R-SWAP \cite{lys2021r}, combined relays and adapters with a hash time lock to tackle safety violations arising from user failures during a swap.
        
        \textbf{Limitations:}
        Atomic swaps implemented using HTLC guarantee loss-proof asset transfers in trustless environments by utilizing decentralization and atomicity \cite{zie2019extending}. However, atomic swaps require both chains to have smart contracts that use HTLC, which is not always possible. Moreover, they can be affected by price speculation during the waiting period of the transaction, such as front running \cite{daian2020flash}.

        Moreover, users have privacy concerns about most atomic protocols based on HTLC, because transactions on both chains are easily identifiable due to the use of the same hash value. In light of this, Deshpande et al. \cite{deshpande2020privacy} proposed an Atomic Release of Secrets (ARS) scheme using Adapter Signatures based on the Schnorr signatures algorithm \cite{neven2009hash} to build cross-chain swaps preserving user privacy.
        
        % Additional discussion and references can be inserted here.

    \subsection{Adaptor Signature-Based Atomic Swaps}
    Adaptor signatures have emerged as a promising cryptographic primitive for enhancing the security and privacy of atomic swap protocols. Several recent works have explored their application in both two-party and multi-party atomic swap scenarios.

    In the context of two-party atomic swaps, Deshpande et al. \cite{deshpande2020privacy} proposed an Atomic Release of Secrets (ARS) scheme based on adaptor signatures and the Schnorr signature algorithm. This approach enhances user privacy by obfuscating transaction details on the blockchain. However, it is limited to two-party swaps and does not address the complexities of multi-party scenarios.
    
    Klamti et al.\cite{klamti2022post} further explored adaptor signatures in the context of atomic swaps, proposing a new two-party adaptor signature scheme that relies on quantum-safe hard problems in coding theory. Their scheme offers enhanced security against quantum computing threats but is limited to two-party scenarios and does not consider the challenges of multi-party atomic swaps.

    For multi-party atomic swaps, Kajita et al. \cite{kajita2024generalized} introduced a generalized adaptor signature scheme for N parties, enabling secure multi-party transactions. Chen et al. \cite{chen2024privacy} proposed a privacy-preserving multi-party cross-chain transaction protocol based on a novel pre-adaptor signature scheme. Ji et al. \cite{ji2023threshold} introduced the concept of threshold adaptor signatures for enhancing the security and fault tolerance of multi-party atomic swaps.

    A comprehensive framework for universal atomic swaps, supporting the secure exchange of assets across multiple blockchains, was presented by Kamp et al. \cite{thyagarajan2022universal}. Their protocol leverages adaptor signatures and time-lock puzzles to achieve practicality and efficiency, but potential vulnerabilities related to time-lock puzzle mechanisms and the computational overhead associated with multiple blockchains remain to be addressed.

    \textbf{Limitations:} While these works have made significant strides in leveraging adaptor signatures for atomic swaps, they often focus on specific scenarios or address a limited set of potential attack vectors. Our proposed protocol aims to overcome these limitations by providing a comprehensive solution for secure, efficient, and scalable multi-party atomic swaps that address a broader range of security concerns.

    \subsection{Sidechains and Wrapped Tokens}
        Sidechains \cite{back2014enabling} are independent blockchains connected to a mainchain and a key to blockchain interoperability. The sidechain approach for PoS blockchains proposed by Gazi et al. involves a two-way peg (2WP) for asset transfers and a merged-staking mechanism for added security during bootstrapping \cite{gavzi2019proof}. RootStock (RSK) provides a federated two-way peg sidechain for Bitcoin and employs a merge-mining scheme for security during the bootstrapping stage \cite{lerner2015rsk}. Additionally, Zendoo uses zk-SNARK for certificate authentication and verification in its cross-chain protocol for asset transfer \cite{garoffolo2020zendoo}.
    
        As described in such projects as BTC-Relay \cite{btcrelaysite_2023}, Cosmos \cite{kwon2019cosmos}, and Polkadot \cite{wood2016polkadot}, sidechains allow assets to be locked on one blockchain, and the equivalent assets to be created in a separate blockchain via a process of locking and burning. Both one-way and two-way asset transfers can be supported between a main and sidechain. The two-way peg mechanism is used to transfer assets at pre-determined exchange rates bidirectionally. The two-way peg can be implemented as a centralized, decentralized, or simple payment verification (SPV) service \cite{singh2020sidechain}.
        
        Wrapped tokens represent assets backed one-to-one by the corresponding asset or collateral held on another chain \cite{9922019}. Wrapped tokens provide a way to move coins and tokens between blockchains that cannot otherwise interoperate. One of the most well-known examples of this concept is Wrapped Bitcoin (WBTC), an ERC-20 token backed 1:1 with Bitcoin. This concept was first introduced by BitGo, Ren, and Kyber Network in a joint initiative in January 2019 \cite{chan2019custodyandfull}.

        Two-way peg mechanisms can be centralized and federated. In the centralized two-way peg, a trusted third party performs token locking, which provides speed and simplicity at the cost of a single point of failure and centralization \cite{singh2020sidechain}. The federated two-way peg distributes control to a group of notaries, reducing centralization and single-point failure issues \cite{back2014enabling,dilley2016strong}.

        \textbf{Limitations:}
        Sidechains share a state with the mainchain, which allows securely locking a token in one chain while it is being used in the other chain. This enables sidechains to  perform instant transactions at a higher speed and volume \cite{9756564}. However, frequent token exchanges between the mainchain and sidechains increase the risks of fraudulent transfers, increase the complexity of interfaces, and also increase the centralization of mining resources \cite{back2014enabling}. Wrapped tokens, on the other hand, require trust in the entity that holds a collateral, which creates a centralized point of failure. Additionally, token wrapping may not be economically practical due to several reasons. First, the implementations must ensure that the value of the wrapped token is the same as the original token. Second, the locked assets need to be safely kept. Third, the peg is maintained over time. Fourth, the tokens can be redeemed at any time, which can cause liquidity shocks \cite{caldarelli2021wrapping}. 
        
        % Additional discussion and references can be inserted here.

    \subsection{Notary-based Token Swap Bridges}

        Notary scheme is a simple solution for cross-chain operations, in which a trusted third party or group observes events on one chain and validates them on another chain \cite{buterin2016chain,wang2021sok}. Variants of this scheme include decentralized and centralized notary mechanisms. Tian et al. \cite{tian2021enabling} devised a cryptocurrency exchange protocol using a decentralized notary scheme to execute atomic swaps. The protocol includes a validation committee (a set of notaries) that checks and validates transactions. Notary election mechanisms mitigate single-point failures. Similarly, RenVM \cite{ren2023interoperability} employs a Byzantine fault-tolerant network coupled with secure multi-party computation (SMPC) to enable sending assets across different blockchains, replacing a trusted centralized custodian with a decentralized, trustless one. Bifrost \cite{scheid2019bifrost} and ReviewChain \cite{wang2018reviewchain} are other examples of notary schemes enabling cross-chain operations. Notary schemes are used by centralized and decentralized exchanges \cite{haugum2022security}.

        \textbf{Limitations:}
            Notary schemes utilize trusted entities (notaries) to verify and sign off on cross-chain transactions. This approach can efficiently handle double-spending and collusion threats. However, notary schemes fundamentally require trust in an external party, compromising blockchain's decentralized and trustless principles. Moreover, they lack adequate resistance to MEV attacks.

    \subsection{Relay-based Token Swap Bridges}

        Relays are used in numerous interoperability protocols. XCLAIM \cite{zamyatin2019xclaim} leverages the BTCRelay protocol to facilitate trustless cross-chain atomic swaps between Bitcoin and Ethereum. It introduces the concept of cryptocurrency-backed assets and includes several entities for complex asset exchange and redeeming requests.
        
        Verilay \cite{westerkamp2022verilay} is the first relay scheme for Proof-of-Stake (PoS) blockchains and is implemented on Ethereum 2.0 \cite{buterin2020combining}. This relay scheme can validate a PoS protocol bys producing the finalized blocks, providing two options for accessing the public keys of validators. Tesseract \cite{bentov2019tesseract} uses a Trusted Execution Environment (TEE) as a trusted relay for secure real-time cryptocurrency exchange. It supports cross-chain trades and asset tokenization schemes that can peg cryptocurrency from one chain to assets on another chain.
        
        Practical AgentChain \cite{hei2022practical} offers a cross-chain exchange providing four features: compatibility, flexibility, high reliability, and strong practicality. It uses smart contracts to resist various attacks and uses Town Crier \cite{zhang2016town} to transmit data from reliable sources. IvyCross \cite{cryptoeprint:2021/1244} is a TEE-based \footnote{TEE stands for "Trusted Execution Environment"} blockchain interoperability framework that offers low-cost, privacy-preserving, and race-free cross-chain communications.

        \textbf{Limitations:}
            Relay schemes enable one blockchain to read and verify transactions from another, providing a certain level of decentralization and atomicity. They offer resistance against double-spending but do not inherently protect against MEV attacks, and do not provide transaction privacy.

    \subsection{Burn-and-Mint Style Protocol}

        In the realm of blockchain interoperability, one notable burn-and-mint style protocol is the Burn-to-Claim protocol as proposed by Pillai, Biswas, Hou, and Muthukkumarasamy \cite{PILLAI2021108495}. This protocol facilitates asset transfers between blockchains by employing a two-step process, which first locks and then subsequently burns the assets on the source blockchain. An equivalent value is then recreated on the destination blockchain.
        
        The Burn-to-Claim protocol proves robust against double-spending attacks, as assets are burnt before they can be claimed on the other chain. It also maintains trustlessness, as no third party is required to oversee the transaction. However, the authors did not provide a proof whether the protocol is inherently resistant to maximum extractable value (MEV) attacks. Malicious miners could potentially manipulate the order of transactions within a block to their advantage, thus compromising the fairness of asset exchange rates.
        
        \textbf{Limitations:}

        While the Burn-to-Claim protocol facilitates asset transfers between blockchains, it exposes transaction details publicly and involves a complex reclaim process if the transfer fails. Another issue is that the current designs of burn-and-mint-style protocols rely on centralized gateways or APIs, which compromise the decentralized trust and security.

    % We use table \ref{tab:compare-bridges} to compare how other protocols differ from our burn-and-mint protocol design.

\section{Multi-Party Atomic Swap Protocol Design}

\subsection{Core Assumptions}
\begin{enumerate}
    \item All participating blockchains support Schnorr-like signature verification.
    \item A secure, decentralized cryptographic accumulator infrastructure is available.
    \item Participants are pre-matched with agreed trading terms.
    \item Effective asset locking mechanisms are implemented.
\end{enumerate}

\subsection{Two-Party Adaptor Signature-Based Atomic Swap Setup}
\subsubsection{Key Generation}
The first step is the same as with a regular Schnorr Signature. Each participant generates their private and public keys by sampling a random value from \( \mathbb{Z}_q \). 

For Alice and Bob:
\[
\begin{aligned}
    & a \leftarrow \mathbb{Z}_q, \quad A = aG \\
    & b \leftarrow \mathbb{Z}_q, \quad B = bG
\end{aligned}
\]

Additionally, both participants generate their random secret nonce values and compute their public derivatives:
\[
\begin{aligned}
    & r_1 \leftarrow \mathbb{Z}_q, \quad R_1 = r_1G \\
    & r_2 \leftarrow \mathbb{Z}_q, \quad R_2 = r_2G
\end{aligned}
\]

Alice and Bob then share their public keys \( A \) and \( B \) as well as their public values \( R_1 \) and \( R_2 \) with each other.

\subsubsection{Adaptor Generation (Alice)}
Alice wants to send Bob a signature over a transaction \( m_1 \) that commits her to transfer the correct amount to him. In a regular Schnorr Signature, Alice would compute:
\[
\begin{aligned}
    & c_1 = H(R_1 \parallel A \parallel m_1) \\
    & s_1 = r_1 + c_1a
\end{aligned}
\]
However, this is a valid signature that can be broadcasted immediately. To prevent Bob from broadcasting it prematurely, Alice modifies the signature by generating a secret value \( t \leftarrow \mathbb{Z}_q \) and its public value \( T = tG \).

Alice then computes:
\[
\begin{aligned}
    & c_1 = H(R_1 + T \parallel A \parallel m_1) \\
    & s_1' = r_1 + t + c_1a
\end{aligned}
\]
She shares \( (s_1', T) \) with Bob.

\subsubsection{Adaptor Verification (Bob)}
Bob verifies that Alice generated a modified signature by checking:
\[
\begin{aligned}
    & c_1 = H(R_1 + T \parallel A \parallel m_1) \\
    & s_1'G \stackrel{?}{=} R_1 + T + c_1A
\end{aligned}
\]

\subsubsection{Adaptor Generation (Bob)}
Bob, having verified Alice's adaptor signature, creates his own. Bob generates his secret value and computes his commitment and partial signature:
\[
\begin{aligned}
    & c_2 = H(R_2 + T \parallel B \parallel m_2) \\
    & s_2 = r_2 + c_2b
\end{aligned}
\]
Bob shares \( s_2 \) with Alice.

\subsubsection{Signature Generation (Alice)}
For Alice to get paid, she uses Bob's adapted signature and her secret value \( t \) to turn it into a valid signature:
\[
\begin{aligned}
    & (s_2 + t)G = R_2 + T + c_2B
\end{aligned}
\]
Once the signature is validated on-chain, Bob can learn \( t \).

\subsubsection{Signature Generation (Bob)}
By observing the blockchain and learning Alice's published signature \( s_2 + t \), Bob can extract the secret value \( t \) and update the adapted signature:
\[
\begin{aligned}
    & t = (s_2 + t) - s_2
\end{aligned}
\]
Bob can now finalize his transaction and get paid as well.

% Three Party
\subsection{Three-Party Atomic Swap Protocol}

\subsubsection{Problem Statement}
We focus on designing a unique atomic swap protocol that operates fully off-chain with only one on-chain finalization. Our goal is to allow more than two parties to securely swap any digital assets on blockchains that support Schnorr signature verification functionality. The main challenge addressed is the malicious dropout attack, common in traditional atomic swap protocols, including those using adaptor signatures.

For example, when Alice, Bob, and Carol decide to trade, Alice sends the pre-signed signature to Bob, who then sends the pre-signed signature to Carol. Carol signs and sends the full signature to Alice to withdraw from Carol. If Alice withdraws money, Carol can maliciously drop out, preventing Bob from seeing the secret to withdraw from Carol. Despite Carol losing money, the atomicity guarantee is broken. Our protocol uses a universal adaptor signature secret to prevent this, ensuring that if Alice can withdraw from the first participant sharing the full signature, all participants can complete their transactions. The protocol scales to any number of participants and operates fully off-chain.

\subsubsection{Setup}
\begin{enumerate}
    \item We set up a secure accumulator accessible to all parties for adding, removing, and checking elements.
    \item The setup results in an environment similar to ECDSA adaptor signature setups for atomic swaps.
\end{enumerate}

\subsubsection{Steps}
1. Instead of only Alice and Bob swapping, we add Carol, making it a 3-party swap. Each participant generates their private keys and nonces:
\[
\begin{aligned}
    & r_3 \leftarrow \mathbb{Z}_p, \quad R_3 = r_3G, \quad c \leftarrow \mathbb{Z}_p, \quad C = cG \\
    & r_1 \leftarrow \mathbb{Z}_p, \quad R_1 = r_1G, \quad a \leftarrow \mathbb{Z}_p, \quad A = aG \\
    & r_2 \leftarrow \mathbb{Z}_p, \quad R_2 = r_2G, \quad b \leftarrow \mathbb{Z}_p, \quad B = bG
\end{aligned}
\]

2. Alice computes her commitment and partial signature:
\[
\begin{aligned}
    & c_A = H(T + R_1 \parallel \text{Acc}(A \parallel B \parallel C) \parallel \text{Acc}(m_1 \parallel m_2 \parallel m_3) \parallel m_1) \\
    & s_A' = r_1 + c_Aa
\end{aligned}
\]
She shares \( (s_A', T) \) with Bob.

3. Bob verifies the commitment and message in the pre-signature, and if they match, he computes his commitment and partial signature:
\[
\begin{aligned}
    & c_B = H(T + R_2 \parallel \text{Acc}(A \parallel B \parallel C) \parallel \text{Acc}(m_1 \parallel m_2 \parallel m_3) \parallel m_2) \\
    & s_B' = r_2 + c_Bb
\end{aligned}
\]
Bob shares this pre-signature \( (c_B, s_B') \) with Carol.

4. Carol verifies the pre-signatures and computes her commitment and partial signature:
\[
\begin{aligned}
    & c_C = H(T + R_3 \parallel \text{Acc}(A \parallel B \parallel C) \parallel \text{Acc}(m_1 \parallel m_2 \parallel m_3) \parallel m_3) \\
    & s_C' = r_3 + c_Cc
\end{aligned}
\]
Carol shares this \( (c_C, s_C') \) with Alice.

5. Once Alice confirms all messages are correct, she computes the full signature:
\[
s_A = s_A' + t
\]
She broadcasts this to her blockchain. All participants can now access their assets on their corresponding blockchains.

\subsubsection{Potential Attacks and Mitigations}
\begin{enumerate}
    \item \textbf{Malicious Dropout:} If Carol drops out after Alice completes her pre-signature, Bob will not get paid, breaking the atomicity promise. To avoid this, our design ensures one universal adaptor secret is shared by the initiator (Alice). Miners and validators must verify adaptor signatures are publicly available and stored in the accumulator. Once Alice completes her adaptor signature, all parties can find the secret \( t \) using \( s_n - s_n' = t \).
    \item \textbf{Skipping Attack:} If Alice skips Bob and directly sends the pre-signature to Carol, Carol will reject the transaction as she cannot receive any asset without Bob's pre-signature.
    \item \textbf{Impersonation Attack:} An attacker (Eve) cannot impersonate a party (Bob) as blockchain signature verification checks public keys in the public key accumulators.
    \item \textbf{Adaptor Signature Leak:} Leaking the adaptor signature does not favor Alice as she must wait until the last party joins to release the secret, making it a non-beneficial action.
\end{enumerate}

\subsubsection{Blockchain Signature Verification Update}
We assume participating blockchains support Schnorr-like signatures. Verification involves checking the validity of signatures using:
\[
c = H(T + R_i \parallel \text{Acc}(\text{pk}_1 \parallel \text{pk}_2 \parallel \ldots) \parallel \text{Acc}(m_1 \parallel m_2 \parallel \ldots))
\]
An accumulator must store all pre-signatures, and full signatures are accepted only if the pre-signature membership proof is verifiable.

\subsubsection{Efficiency}
Using the accumulator and off-chain setups, the extra on-chain overhead is minimal, as all swap transactions are off-chain, requiring only the final on-chain finalization step.

% N-Party
\subsection{N-Party Atomic Swap}

Our protocol is designed to generalize efficiently from a multi-party atomic swap to an N-party atomic swap, where \(N\) represents the total number of participants involved in the swap. This section provides a detailed explanation of how the protocol can be extended to handle \(N\) participants, including the logic and mathematical formulations required to maintain security, atomicity, and efficiency.

\subsubsection{Generalized Setup for N-Party Swaps}

In an N-party atomic swap, each participant \( P_i \) (where \( i \in \{1, 2, \ldots, N\} \)) generates their private and public keys, nonce values, and commitments as follows:

For each participant \( P_i \):
\[
\begin{aligned}
  & a_i \leftarrow \mathbb{Z}_q, \quad A_i = a_iG \\
  & r_i \leftarrow \mathbb{Z}_q, \quad R_i = r_iG
\end{aligned}
\]

Each participant shares their public keys \( A_i \) and nonce values \( R_i \) with all other participants. Additionally, the first participant \( P_1 \) generates a secret value \( t \leftarrow \mathbb{Z}_q \) and its public value \( T = tG \), which is shared with all participants.

\subsubsection{Generalized Adaptor Generation}

To initiate the swap, the first participant \( P_1 \) computes the commitment and partial signature as follows:
\[
\begin{aligned}
  & \text{Acc}_{A} = \text{Acc}(A_1 \parallel A_2 \parallel \ldots \parallel A_N) \\
  & \text{Acc}_{m} = \text{Acc}(m_1 \parallel m_2 \parallel \ldots \parallel m_N) \\
  & c_1 = H\left(R_1 + T \parallel A_1 \parallel \text{Acc}_{A} \parallel \text{Acc}_{m} \parallel m_1\right) \\
  & s_1' = r_1 + t + c_1a_1
\end{aligned}
\]
\( P_1 \) then shares \( (s_1', T) \) with the next participant \( P_2 \).

\subsubsection{Verification and Adaptor Generation for Subsequent Participants}

Each subsequent participant \( P_i \) (where \( i \in \{2, 3, \ldots, N-1\} \)) verifies the previous participant's commitment and partial signature, then computes their own commitment and partial signature as follows:
\[
\begin{aligned}
  & c_i = H\left(R_i + T \parallel A_i \parallel \text{Acc}_{A} \parallel \text{Acc}_{m} \parallel m_i\right) \\
  & s_i' = r_i + c_i a_i
\end{aligned}
\]
\( P_i \) then shares \( (c_i, s_i', T) \) with the next participant \( P_{i+1} \).

\subsubsection{Finalization by the Last Participant}

The final participant \( P_N \) verifies all previous participants' commitments and partial signatures, then computes their own commitment and full signature as follows:
\[
\begin{aligned}
  & c_N = H\left(R_N + T \parallel A_N \parallel \text{Acc}_{A} \parallel \text{Acc}_{m} \parallel m_N\right) \\
  & s_N = r_N + c_N a_N
\end{aligned}
\]
\( P_N \) then broadcasts the full signature \( s_N \) to the blockchain. 

\subsubsection{Completion and Asset Retrieval}

Upon seeing \( s_N \) on the blockchain, each participant \( P_i \) can extract the secret \( t \) as follows:
\[
t = s_N - s_N'
\]
With the secret \( t \), each participant can compute their full signature and retrieve their assets. For example, participant \( P_i \) computes:
\[
s_i = s_i' + t
\]

\subsubsection{Implications and Security Analysis}

The generalized protocol ensures atomicity and security for \(N\)-party swaps by maintaining the dependency on a single adaptor secret \( t \) and cryptographic accumulators. Each participant's ability to compute their full signature relies on the integrity of the previous participants' commitments, making it impossible to skip any participant or impersonate another.

This structure also ensures that even if one participant attempts a malicious dropout, the swap can still be completed by the other participants, preserving atomicity. The use of cryptographic accumulators provides an additional layer of security, ensuring that only legitimate participants with valid public keys can complete their transactions.

By extending the protocol to \(N\) participants, we maintain the same level of efficiency and security, with minimal on-chain overhead. This makes our protocol highly scalable and practical for real-world blockchain applications, supporting secure and efficient cross-chain asset exchanges among multiple participants.

\section{Security Analysis}

    In this section, we present a security analysis of our proposed multi-party atomic swap protocol. We identify potential attack vectors and demonstrate how our protocol mitigates these risks through its design and the utilization of cryptographic primitives.
    
    \subsection{Malicious Dropout Attack}
    
    The malicious dropout attack is a significant concern in multi-party atomic swaps, where a participant may attempt to disrupt the atomicity of the swap by refusing to complete their part of the transaction after receiving assets from another participant. In our protocol, the universal adaptor secret, denoted as (t), plays a crucial role in preventing this attack.
    
    Let's consider a scenario where Alice initiates a swap with Bob and Carol. Alice generates a pre-signature $(s_A', T)$ and shares it with Bob. Bob, in turn, generates his pre-signature $(c_B, s_B')$ and shares it with Carol. Carol then generates $(c_C, s_C')$ and shares it with Alice.
    
    If Carol attempts a dropout attack after Alice finalizes her transaction by broadcasting $(s_A = s_A' + t)$, Bob can still compute the secret (t) using the publicly available pre-signatures:
    
    $$
    t = s_C - s_C'
    $$
    
    With the knowledge of (t), Bob can compute his full signature $(s_B = s_B' + t)$ and finalize his transaction, ensuring that he receives his share of the assets. This mechanism guarantees atomicity even in the presence of malicious dropouts, as any participant can complete their transaction if the initiator finalizes theirs.
    
    \subsection{Skipping Attack}
    
    In a skipping attack, a participant attempts to bypass an intermediate participant and directly interact with a subsequent participant in the swap chain. For instance, Alice might try to send her pre-signature directly to Carol, bypassing Bob. However, our protocol's design inherently prevents this attack.
    
    Carol, upon receiving Alice's pre-signature, would be unable to compute her full signature $(s_C)$ as she requires Bob's pre-signature $(c_B, s_B')$ to compute her commitment $(c_C)$. This dependency on the pre-signatures of preceding participants ensures that each participant must follow the designated swap chain, preventing any attempts to skip intermediate participants.
    
    \subsection{Impersonation Attack}
    
    An impersonation attack involves a malicious actor attempting to impersonate a legitimate participant in the atomic swap. Our protocol mitigates this risk through the use of cryptographic accumulators and public key verification.
    
    Each participant's public key is included in a public key accumulator, $(\text{Acc}(A|B|C)$, which is used in the computation of commitments and signatures. When a participant broadcasts a transaction, the blockchain verifies the validity of the signature and checks whether the sender's public key is a member of the accumulator. This ensures that only legitimate participants can finalize transactions, preventing unauthorized access to assets.
    
    \subsection{Adaptor Signature Leakage}
    
    While the leakage of adaptor signatures could potentially disrupt the atomicity of traditional atomic swaps, our protocol's design ensures that such leakage does not compromise the security or fairness of the swap.
    
    Even if Alice, the initiator, were to leak her adaptor signature $(s_A', T)$, it would not provide any advantage to the attacker. The attacker would still need the final participant's full signature to extract the secret (t) and complete their own transaction. Furthermore, the public nature of pre-signatures ensures that all participants have access to the necessary information to complete their transactions, regardless of any leakage.
    
    Our proposed multi-party atomic swap protocol offers robust security guarantees against various attack vectors, ensuring the atomicity, fairness, and security of cross-chain asset exchanges. The combination of universal adaptor secrets, cryptographic accumulators, and public key verification creates a secure and trustless environment for multi-party atomic swaps.
\section{Discussion}

The security analysis presented in the previous section demonstrates the robustness of our proposed multi-party atomic swap protocol against various attack vectors. However, it's crucial to acknowledge the specific conditions and assumptions under which these security guarantees hold.

Reliance on Secure Accumulator Infrastructure: Our protocol's security heavily relies on the availability of a secure and trustworthy accumulator infrastructure. This infrastructure is responsible for storing and managing pre-signatures, ensuring their public verifiability, and providing membership proofs for full signature verification. Any compromise of this infrastructure could potentially undermine the security of the entire protocol.

To mitigate this risk, it's essential to employ a decentralized and robust accumulator infrastructure, potentially utilizing a consortium blockchain or a distributed network of trusted nodes. Implementing redundancy and fault tolerance mechanisms within the infrastructure can further enhance its security and resilience.

Assumption of Pre-matched Participants: Our protocol assumes that participants are pre-matched with agreed-upon trading factors such as time and price. While this assumption simplifies the protocol's design and operation, it may not always hold in real-world scenarios. In decentralized environments, participants may need to discover and negotiate trading terms dynamically.

To address this, future extensions of our protocol could incorporate mechanisms for decentralized matchmaking and negotiation, potentially leveraging smart contracts or decentralized exchange protocols. However, integrating these mechanisms while maintaining the protocol's security and efficiency would require careful consideration and design.

Dependence on Effective Asset Locking: Our protocol assumes the existence of effective asset locking mechanisms to prevent participants from spending assets prematurely. The security of these locking mechanisms is crucial for ensuring the atomicity and fairness of the swap.

Different blockchains may offer varying levels of asset locking capabilities, and the choice of blockchain could impact the overall security of the protocol. It's essential to select blockchains with robust and secure asset locking mechanisms to minimize the risk of premature spending and ensure the protocol's effectiveness.

Potential for Collusion Attacks: While our protocol mitigates several attack vectors, the possibility of collusion attacks, where multiple participants collude to defraud others, cannot be entirely ruled out. For instance, if the majority of participants in a swap collude, they could potentially manipulate the accumulator or other aspects of the protocol to their advantage.

Mitigating collusion attacks in multi-party atomic swaps remains a challenging open problem. Potential solutions could involve incorporating reputation systems, requiring deposits or collateral from participants, or utilizing advanced cryptographic techniques such as threshold signatures or multi-party computation. However, implementing these solutions would likely introduce additional complexity and overhead to the protocol.

Our proposed multi-party atomic swap protocol offers a significant advancement in the field of cross-chain asset exchange. Its focus on off-chain operation, scalability, and security through cryptographic primitives makes it a promising solution for decentralized finance applications. However, it's crucial to acknowledge the limitations and assumptions of the protocol, and future research should explore ways to address these limitations and enhance the protocol's robustness in real-world scenarios.
\section{Conclusion}
\label{sec:conclusion}

    Our proposed protocol for atomic swaps utilizing adaptor signatures and cryptographic accumulators offers a robust and scalable solution for cross-chain transactions. By addressing the limitations of traditional atomic swap protocols, our approach enhances the efficiency and security of cross-chain interactions. Adaptor signatures significantly reduce the computational and storage overhead on blockchains, facilitating the simultaneous preparation and execution of multiple transactions. Cryptographic accumulators enable efficient and verifiable management of large transaction batches with minimal on-chain operations. The protocol's design ensures strong security guarantees, leveraging advanced cryptographic techniques to protect against common attack vectors such as replay attacks and fraudulent proofs. This trustless mechanism aligns with the core principles of decentralization, providing a secure framework for cross-chain transactions without relying on intermediaries.
\bibliographystyle{ACM-Reference-Format}
% \bibliography{src/bibs/main}


\appendix

\section{ECDSA-Based Adaptor Signature for Atomic Swaps}
\label{app:adaptor-signature}
    
    Adaptor signatures \cite{poelstra2017scriptless} extend traditional digital signatures by incorporating a cryptographic condition that can reveal a secret value. This feature is particularly useful in blockchain applications for ensuring fairness in transactions such as atomic swaps \cite{tu2022efficient}\cite{kajita2024generalized}.
    
    \subsection{Notations and Preliminaries}
    
    Let \( \mathbb{G} \) be an elliptic curve group of prime order \( q \) with a generator \( G \). The security parameter is denoted by \( \lambda \). We use \( \text{GenR}(1^\lambda) \) to denote a probabilistic polynomial-time (PPT) algorithm that generates statement/witness pairs \( ((G, Y = yG), y) \) for a hard relation \( R \). The associated language \( L_R \) is defined as \( L_R = \{(G, Y) \mid \exists y \text{ such that } ((G, Y), y) \in R\} \).
    
    A non-interactive zero-knowledge proof of knowledge (NIZKPoK) with straight-line extractors is defined as a pair of PPT algorithms \( (P, V) \) that satisfies completeness, zero-knowledge, and straight-line extractability.
    
    \subsection{ECDSA-Based Adaptor Signature Scheme}
    
    An ECDSA-based adaptor signature scheme $$ \Pi_R = (\text{pSign}, \text{pVrfy}, \text{Adapt}, \text{Ext}) $$ consists of the following algorithms:
    \begin{enumerate}
    \item Pre-signing (\( \text{pSign} \)):
       \[
       \text{pSign}_{sk}(m, Y) \rightarrow \hat{\sigma}
       \]
       On input a signing key \( sk \), an instance \( Y \), and a message \( m \), it outputs a pre-signature \( \hat{\sigma} \).
    \item Pre-signature Verification (\( \text{pVrfy} \)):
       \[
       \text{pVrfy}_{vk}(m, Y, \hat{\sigma}) \rightarrow \{0, 1\}
       \]
       On input a verification key \( vk \), a message \( m \), an instance \( Y \), and a pre-signature \( \hat{\sigma} \), it outputs 1 if the pre-signature is valid, and 0 otherwise.
    \item Adaptor (\( \text{Adapt} \)):
       \[
       \text{Adapt}(\hat{\sigma}, y) \rightarrow \sigma
       \]
       On input a pre-signature \( \hat{\sigma} \) and a witness \( y \), it outputs a full signature \( \sigma \).
    \item Extraction (\( \text{Ext} \)):
       \[
       \text{Ext}(\sigma, \hat{\sigma}, Y) \rightarrow y
       \]
       On input a signature \( \sigma \), a pre-signature \( \hat{\sigma} \), and an instance \( Y \), it extracts the witness \( y \).
    \end{enumerate}
    
    \subsection{Security Definitions}
    \begin{enumerate}
    \item Pre-signature Correctness:
       The scheme satisfies pre-signature correctness if for all \( \lambda \), every message \( m \), and every statement/witness pair \( (Y, y) \in R \), the following holds:
       \[
       \Pr \left[ 
       \begin{array}{c}
       \text{pVrfy}_{vk}(m, Y, \hat{\sigma}) \rightarrow 1 \\
       \text{Vrfy}_{vk}(m, \sigma) \rightarrow 1 \\
       (Y, y') \in R
       \end{array} \right] = 1
       \]
       where \( \hat{\sigma} \leftarrow \text{pSign}_{sk}(m, Y) \), \( \sigma \leftarrow \text{Adapt}(\hat{\sigma}, y) \), and \( y' \leftarrow \text{Ext}(\sigma, \hat{\sigma}, Y) \).
    \item Existential Unforgeability under Chosen Message Attack (aEUF-CMA):
       The scheme is aEUF-CMA secure if for every PPT adversary \( A \), there exists a negligible function \( \text{negl} \) such that:
       \[
       \Pr[\text{aSigForge}_{A, \Pi_R}(\lambda) = 1] \leq \text{negl}(\lambda)
       \]
    \item Pre-signature Adaptability:
       The scheme satisfies pre-signature adaptability if for any \( \lambda \), any message \( m \), any statement/witness pair \( (Y, y) \in R \), any key pair \( (vk, sk) \leftarrow \text{Gen}(1^\lambda) \), and any pre-signature \( \hat{\sigma} \) with \( \text{pVrfy}_{vk}(m, Y, \hat{\sigma}) \rightarrow 1 \), we have \( \text{Vrfy}_{vk}(m, \text{Adapt}(\hat{\sigma}, y)) \rightarrow 1 \).
    \item Witness Extractability:
       The scheme is witness extractable if for every PPT adversary \( A \), there exists a negligible function \( \text{negl} \) such that:
       \[
       \Pr[\text{aWitExt}_{A, \Pi_R}(\lambda) = 1] \leq \text{negl}(\lambda)
       \]
    \end{enumerate}
    
    \subsection{ECDSA-Based Adaptor Signature Construction}
    
    Let \( (Q = xG, x) \) be the verification key and signing key of ECDSA. The scheme \( \Pi_R \) is constructed as follows:
    \begin{enumerate}
    \item Pre-signing:
       \[
       \text{pSign}_{(vk, sk)}(m, I_Y) \rightarrow \hat{\sigma}
       \]
       On input a key pair \( (vk, sk) = (Q, x) \), a message \( m \), and an instance \( I_Y = (Y, \pi_Y) \), the algorithm computes the pre-signing public parameter \( Z = xY \), runs \( \pi_Z \leftarrow P_Z(I_Z = (G, Q, Y, Z), x) \), chooses \( k \leftarrow \mathbb{Z}_q \), computes \( r = f(kY) \), \( \hat{s} = k^{-1}(h(m) + rx) \mod q \), and outputs the pre-signature \( \hat{\sigma} = (r, \hat{s}, Z, \pi_Z) \).
    \item Pre-signature Verification:
       \[
       \text{pVrfy}_{vk}(m, I_Y, \hat{\sigma}) \rightarrow 0/1
       \]
       On input the verification key \( vk = Q \), a message \( m \), an instance \( I_Y \), and a pre-signature value \( \hat{\sigma} \), the algorithm outputs 0 if \( V_Z(I_Z) \rightarrow 0 \); otherwise, it computes \( r' = f(\hat{s}^{-1}(h(m)Y + rZ)) \mod q \), and if \( r' = r \), it outputs 1; else, it outputs 0.
    \item Adaptor:
       \[
       \text{Adapt}(y, \hat{\sigma}) \rightarrow \sigma
       \]
       On input the witness \( y \) and pre-signature \( \hat{\sigma} \), the algorithm computes \( s = \hat{s} \cdot y^{-1} \mod q \) and outputs the signature \( \sigma = (r, s) \).
    \item Extraction:
       \[
       \text{Ext}(\sigma, \hat{\sigma}, I_Y) \rightarrow y
       \]
       On input the signature \( \sigma \), the pre-signature \( \hat{\sigma} \), and the instance \( I_Y \), it computes \( y = \hat{s} / s \mod q \). If \( (I_Y, y) \in R \), it outputs \( y \); else, it outputs \( \perp \).
    \end{enumerate}
    
    \subsection{Security Anaysis}
    
    The security of the ECDSA-based adaptor signature scheme relies on the strong unforgeability of positive ECDSA and the properties of the NIZKPoK. We outline the security analysis for pre-signature adaptability, pre-signature correctness, aEUF-CMA security, and witness extractability.
    \begin{enumerate}
    \item Pre-signature Adaptability:
       Let \( (I_Y, y) \in R \), \( m \in \{0, 1\}^* \), and \( \hat{\sigma} = (r, \hat{s}, Z, \pi_Z) \) be a valid pre-signature with \( \text{pVrfy}_{vk}(m, I_Y, \hat{\sigma}) \rightarrow 1 \). By definition of the adaptor algorithm, \( \sigma = (r, s) \) where \( s = \hat{s} \cdot y^{-1} \mod q \). We have:
       \[
       K' = (h(m) \cdot s^{-1})G + r \cdot s^{-1}Q = kY
       \]
       Thus, \( r' = f(K') = r \), implying \( \text{Vrfy}_{vk}(m, \sigma) \rightarrow 1 \).
    \item Pre-signature Correctness:
       For any \( x, y \in \mathbb{Z}_q \), let \( Q = xG \), \( Y = yG \), and \( m \in \{0, 1\}^* \). For \( \hat{\sigma} = (r, \hat{s}, Z, \pi_Z) \) such that \( \hat{\sigma} \leftarrow \text{pSign}_{vk, sk}(m, I_Y) \), we have:
       \[
       Y = yG, \quad Z = xY, \quad \hat{s} = k^{-1}(h(m) + rx) \mod q \text{ for some } k \leftarrow \mathbb{Z}_q
       \]
       Setting \( K' = (h(m) \cdot \hat{s}^{-1})Y + r \cdot \hat{s}^{-1}Z = kY \), we obtain \( r' = f(K') = f(kY) = r \). Thus, \( \text{pVrfy}_{vk}(m, I_Y, \hat{\sigma}) \rightarrow 1 \). This implies \( \text{Vrfy}_{vk}(m, \sigma) \rightarrow 1 \) for \( \sigma = (r, s) \) where \( s = \hat{s} \cdot y^{-1} \).
    \item aEUF-CMA Security:
       Assuming the positive ECDSA is SUF-CMA secure, the hard relation \( R \), NIZKPoK, and NIZK are secure. For any PPT adversary \( A \) breaking aEUF-CMA security of the ECDSA-AS, we construct a PPT simulator \( S \) who breaks the SUF-CMA security of ECDSA. \( S \) uses its signing oracle \( \mathcal{O}_{\text{ECDSA-Sign}} \) and random oracle \( H_{\text{ECDSA}} \) to simulate oracles for \( A \). The simulator extracts the witness from \( I_Y \) and uses the zero-knowledge property of \( \pi_Z \) to simulate proofs without knowing the corresponding witness \( x \).
    \item Witness Extractability:
       The proof reduces witness extractability to the strong unforgeability of positive ECDSA. The simulator \( S \) uses \( \mathcal{O}_{\text{ECDSA-Sign}} \) and \( H_{\text{ECDSA}} \) to simulate oracles for \( A \). The crucial difference between \( \text{aWitExt} \) and \( \text{aSigForge} \) is the choice of \( I_Y \). The simulator extracts the witness \( y \) from \( I_Y \) chosen by \( A \), ensuring the simulator wins the positive ECDSA strongSigForge game if \( A \) can break the witness extractability of ECDSA-AS.
    \end{enumerate}
    
    The ECDSA-based adaptor signature scheme provides a robust and efficient method for performing atomic swaps on the blockchain. The scheme ensures fairness and security in cryptocurrency exchanges by leveraging zero-knowledge proofs and the adaptability of pre-signatures.

\section{Scalable Blockchain State Verification with Accumulator and SNARK}
\label{app:accumulator}

    This section explains the verifiable computation protocol for blockchain using set accumulators inspired by the works of Boneh et al. \cite{ozdemir2020scaling}.
    
    \subsection{Background and Definitions}
    
    \subsubsection{Multisets}
    A multiset is an unordered collection that may contain multiple copies of any element. Let \(S_1, S_2\) denote two multisets. The union of multisets \(S_1\) and \(S_2\) is denoted \(S_1 \cup S_2\), where the multiplicity of each element \(x\) in \(S_1 \cup S_2\) is the sum of the multiplicities of \(x\) in \(S_1\) and \(S_2\). The difference of multisets \(S_1\) and \(S_2\) is denoted \(S_1 \setminus S_2\), where each element \(x\) in \(S_1 \setminus S_2\) has multiplicity equal to the difference of multiplicities of \(x\) in \(S_1\) and \(S_2\).
    
    \subsubsection{RSA Groups}
    An RSA group is the group \(\mathbb{Z}_N^*\), i.e., the multiplicative group of invertible integers modulo \(N\), where \(N\) is the product of two secret primes. The RSA quotient group for \(N\) is defined as the group \(\mathbb{Z}_N^*/\{\pm 1\}\). In this group, elements \(x\) and \(-x\) are considered the same, meaning that all elements can be represented by integers in the interval \([1, N/2]\).
    
    \subsubsection{Proofs and Arguments}
    Informally, a proof is a protocol between a prover \(P\) and a polynomial-time verifier \(V\) by which \(P\) convinces \(V\) that \(\exists w: R(x, w) = 1\) for a relation \(R\), input \(x\) from \(V\), and witness \(w\) from \(P\). A proof satisfies the following properties:
    \begin{itemize}
        \item Completeness: If \(\exists w: R(x, w) = 1\), then an honest \(P\) convinces \(V\) except with probability \(\epsilon_c \ll \frac{1}{2}\).
        \item Soundness: If \(\not\exists w: R(x, w) = 1\), no cheating prover \(P^*\) convinces \(V\) except with probability \(\epsilon_s \ll \frac{1}{2}\).
    \end{itemize}
    
    \subsection{RSA Accumulators}
    An RSA multiset accumulator represents a multiset \(S\) with the digest:
    \[
    [S] = g^{\prod_{s \in S} H(s)} \mod N
    \]
    where \(g\) is a fixed member of an RSA quotient group \(\mathbb{G}\), and \(H\) is a division-intractable hash function. Inserting a new element \(s\) into \(S\) requires computing \([S \cup \{s\}] = [S]^{H(s)} \mod N\).
    
    To prove membership of \(s \in S\), the prover furnishes the value \(\pi = [S]^{1/H(s)} \mod N\), which is verified by checking that \(\pi^{H(s)} = [S] \mod N\).
    
    Non-membership proofs are also possible, leveraging the fact that \(s \notin S\) if and only if \(\gcd(H(s), \prod_{s' \in S} H(s')) = 1\).
    
    \subsection{MultiSwap Operation}
    The MultiSwap operation provides a sequential update semantics for RSA accumulators. It takes an accumulator and a list of pairs of elements, removing the first element from each pair and inserting the second. Formally, given a multiset \(S\) and a sequence of swaps \((x_1, y_1), \ldots, (x_n, y_n)\), MultiSwap produces a new multiset \(S_t\) such that:
    \[
    S_t = S \setminus \{x_1, \ldots, x_n\} \cup \{y_1, \ldots, y_n\}
    \]
    
    \subsubsection{Proofs for Batched Operations}
    The RSA accumulator supports batched insertions through an interactive protocol. The final check for a batch insertion proof is:
    \[
    Q^{\ell} \cdot [S]^{\prod_{i=1}^k H(y_i)} \mod \ell = [S \cup \{y_1, \ldots, y_k\}] \mod N
    \]
    where \(\ell\) is a random prime challenge. Removing elements \(x_1, \ldots, x_k\) from \(S\) is similar:
    \[
    Q^{\ell} \cdot [S \setminus \{x_1, \ldots, x_k\}] \mod \ell = [S]
    \]
    
    \subsection{Hash Functions}
    \subsubsection{Hashing to Primes}
    The hash function \(H_p\) generates the challenge \(\ell\) used in Wesolowski proofs. \(H_p\) must output primes with sufficient entropy, which is typically achieved through rejection sampling and probabilistic primality testing (e.g., Miller-Rabin).
    
    \subsubsection{Division-Intractable Hashing}
    A hash function \(H\) is division-intractable if it is infeasible to find \(x, \{x_i\}\) such that \(H(x)\) divides \(\prod_i H(x_i)\). Coron and Naccache show that a hash function that outputs sufficiently large integers is division-intractable.

\end{document}